\documentclass[prl,letterpaper,twocolumn,showpacs,superscriptaddress,floatfix]{revtex4}

\usepackage{graphicx,psfrag,amsmath,amssymb,amsfonts,bbm,latexsym,color,dcolumn,bm}

\begin{document}

\title{Edge effects in electrostatic calibrations for the measurement of the Casimir force}

\author{Qun Wei}
\affiliation{Department of Physics and Astronomy,Dartmouth College,6127 Wilder Laboratory,Hanover,NH 03755,USA}

\author{Roberto Onofrio}
\affiliation{Dipartimento di Fisica ``Galileo Galilei'',Universit\`a  di Padova,Via Marzolo 8,Padova 35131,Italy}

\affiliation{Department of Physics and Astronomy,Dartmouth College,6127 Wilder Laboratory,Hanover,NH 03755,USA}

\date{\today}

\begin{abstract}
We have performed numerical simulations to evaluate the effect on the
capacitance of finite size boundaries realistically present in the 
parallel plane, sphere-plane, and cylinder-plane geometries. 
The potential impact of edge effects in assessing the accuracy of 
the parameters obtained in the electrostatic calibrations of Casimir 
force experiments is then discussed.  
\end{abstract}


\maketitle

\section{1. Introduction}

\begin{figure*}[t]
\includegraphics[width=0.9\textwidth, clip=true]{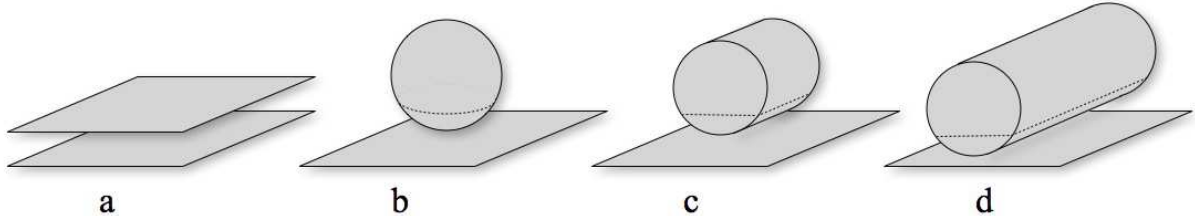}
\caption{Selected geometries for the numerical study of finite size effects. 
Parallel plates of finite size (a); a sphere in front of a finite size plane (b); 
a cylinder in front of a plane, with width smaller (c) and larger (d) than the size 
of the plane. In the latter three configurations, a truncated
sphere and short and long truncated cylinders, as shown by the dashed lines, have been also studied.}
\label{fig:comsol_shape}
\end{figure*}

The Casimir force \cite{Casimir} has been demonstrated in a variety of
experimental setups and geometries, yet there is an ongoing reanalysis 
of the level of accuracy with which it has been determined, which is
crucial for assessing reliable limits to the existence of Yukawa forces of 
gravitational origin predicted by various models \cite{Fishbach,Reynaudrev,Onofrio}. 
In performing Casimir force measurements a crucial role is played by 
the related electrostatic force calibrations, and mastering all
possible systematic effects in the latter is mandatory to assess the 
precision of the former. 
The presence of anomalous exponents in the power-law dependence upon
distance and the dependence on distance of the minimizing potential 
\cite{PRARC} have been identified as a possible source of systematic 
effects, to be carefully scrutinized in each experimental setup \cite{Group}.
The distance-dependence of the minimizing potential has been recently 
modelized theoretically, after a first effort reported in
\cite{Speake}, in terms of non-equipotential conducting surfaces 
due to random patterns of patch charges \cite{KimPRA}.  
Here we report results on a further potential source of systematic
error by studying, by means of numerical simulations using finite
element analysis, the influence of the unavoidable presence of edges on electrostatic 
calibrations in various geometries (for the influence of edge effects on Casimir 
forces in peculiar geometries see \cite{Gies1,Gies2}). 
We limit the attention to the three geometries that, apart from the 
crossed-plane investigated in \cite{Ederth}, have been extensively 
discussed for measuring Casimir force so far, {\it i.e.} the parallel plates 
\cite{Spaarnay,Bressi}, the sphere-plane \cite{Lamoreaux,Mohideen,Chan,Decca}, 
and the recently proposed cylinder-plane \cite{Michael1} (see Fig.~\ref{fig:comsol_shape}). 
We discuss the deviations from ideality in all these geometries
through the essential knowledge of the capacitance dependence on
distance, and its general implications for determining the parameters
used in the measurement of the Casimir force. This is performed under
the simplyfing assumptions that the surfaces are equipotential, 
{\it i.e.} by omitting any superimposed effect due to electrostatic
patches, and by neglecting the tensorial nature of the capacitance among
all the conducting surfaces realistically involved in concrete 
experimental setups.

\section{2. Parallel plate geometry}

\begin{figure*}[t]
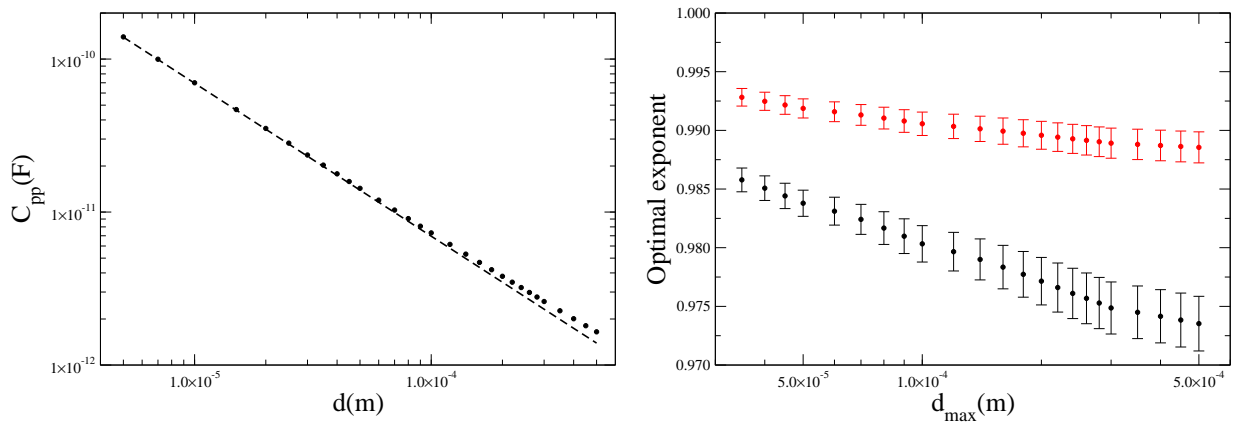

\includegraphics[width=0.45\textwidth, clip=true]{edge.fig2a.eps}
\includegraphics[width=0.45\textwidth, clip=true]{edge.fig2b.eps}
\caption{(Left) Capacitance {\it vs.} distance for two parallel square plates. 
Dots indicate the results of the numerical simulation, while the dashed line 
is the capacitance expected using the formula $C_\mathrm{pp} = \epsilon_0 A/d$. 
The length of the square plates $L$ was chosen to be 8.86 mm.
(Right) Optimal exponent $\epsilon_\mathrm{c}$ coming from the best fit of 
a set of numerical data with $C_\mathrm{pp} = \epsilon_0 A/(d-d_0)^{\epsilon_\mathrm{c}}$
{\it vs.} the distance of the farthest data point used in the fitting. 
Red dots are obtained by fixing $d_0$=0, corresponding to an {\it a priori} 
knowledge of the absolute distance, while the black squares are obtained if 
$d_0$ is considered as a fitting parameter, as usually done when analyzing 
real experimental data with no {\it a priori} independent knowledge of the
absolute distance between the two surfaces. Considering the size of
the square plates used in the experiment reported in \cite{Bressi} of
1.1 mm, all separation distances in that case are obtained by scaling down 
the horizontal axis by a factor $\simeq 65$.}
\label{fig:comsol_ppsquare}
\end{figure*}

We start our analysis with the parallel plates configuration since this is the simplest 
geometry even in terms of possible deviations from the ideal, infinite
plane case. Moreover, besides having an exact expression in the ideal
case of infinite plates, analytical approximate expressions are also 
available for the capacitance including edge effects \cite{Wintle},
allowing to obtain reliable numerical benchmarks.
We consider two identical parallel square plates of length $L$, and
boundary conditions set in such a way that the two plates are at 
a constant electric potential difference. 
The total electrostatic potential energy is then computed by
numerically solving the Laplace equation using a dedicated 
finite element analysis software (COMSOL). 
We evaluate the total electrostatic energy $W_\mathrm{el}$ by summing
the electrostatic energy density over a selected volume surrounding 
the two equipotential surfaces.
The capacitance $C$ can then be calculated from the relationship $C = 2 W_\mathrm{el} V^2$. 
The parameters of the mesh (mesh size, rate of growth etc.) are 
carefully chosen and extensively tested to ensure the accuracy of the numerical 
results. When the distance between the two plates is much smaller than $L$, 
the capacitance obtained from the numerical simulation agrees within 0.01$\%$ with 
the well known analytical formula $C_\mathrm{pp} = \epsilon_0 A/d$, 
where $A$ is the surface area of the plates, $d$ their separation, and $\epsilon_0$ 
the vacuum electric permittivity. However, when the distance becomes larger, 
the value of the capacitance begins to deviate from the analytical formula.
The left plot in Fig.~\ref{fig:comsol_ppsquare} shows the capacitance
of two parallel square plates {\it vs.} distance, together with the 
expected capacitance for two plates following the infinite 
surface formula, appearing as a dashed line. The data at large distance deviate 
significantly from the dashed line, and the fact that they are all
above implies that the power-law exponent is softer, {\it i.e.} in between 0 and -1. 
This deviation is ascribed to the finite size of the parallel plates. 
In fact, if only the field lines in the volume delimited by the two plates are used
for the sum of the electrostatic energy, which corresponds to ignoring
the contribution of the outer region, the capacitance obtained from the numerical simulation would
still agree with the analytical formula even at the largest explored distances. 
In order to better quantify this deviation we have fitted the capacitance curve by 
progressively removing points at the largest distance. 
As shown in the right plot in Fig.~\ref{fig:comsol_ppsquare}, the
optimal exponent becomes smaller than unity when data at large
distance are progressively included in the fit. 

\section{3. Sphere-plane geometry}

\begin{figure}[t]
\includegraphics[width=0.45\textwidth, clip=true]{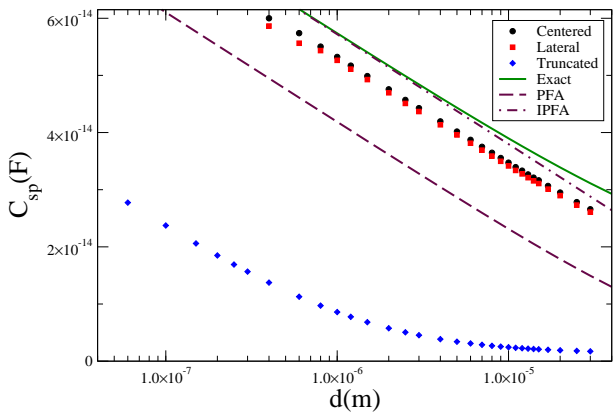}
\caption{Capacitance dependence on distance in the sphere-plane configurations. 
The case of a whole sphere located above the center of a square plate (black dots) 
and on the edge of the plate (red squares) are shown, as well as the case of a 
truncated sphere (blue diamonds). The radius of the sphere is 
chosen to be 0.15 mm and the length of the square plate 0.5 mm. 
The continuous line corresponds to the exact expression as in (\ref{eq:sphereexact}), 
and the dashed dot line corresponds to the approximate expression as in 
(\ref{eq:sphereapprox}), named Improved PFA (IPFA), while the dashed line 
is the capacitance expected from the PFA method.} 
\label{fig:comsol_sp}
\end{figure}

In the case of the sphere-plane geometry, the exact expression for the capacitance 
between a sphere and an infinite plane is written in terms of a series
\cite{Boyer}:
\begin{equation}
C_\mathrm{sp} = 4 \pi \epsilon_0 R \sinh(\alpha) \sum_{n=1}^{+\infty} \frac{1}{\sinh(n \alpha)}
\label{eq:sphereexact}
\end{equation}
with $\cosh(\alpha) = 1 + d/R$, $R$ is the sphere radius, and $d$ the separation distance.
In the limit of small separations, $d/R << 1$, an approximate expression for 
the capacitance can be obtained \cite{Boyer}:
\begin{equation}
C_\mathrm{sp} \approx 2 \pi \epsilon_0 R \left(\ln \frac{R}{d} + \ln 2 + \frac{23}{20} + \frac{\theta}{63}\right)
\label{eq:sphereapprox}
\end{equation}
where $0 \leq \theta \leq 1$. By neglecting the distance-independent terms one obtains 
an expression often appearing in relationship to the so-called Proximity Force Approximation (PFA) 
\cite{Derjaguin,Blocki} in electrostatics. 
The formulas above both assume an infinite plane and a whole sphere. 
In real experiments involving microresonators the size
of the plane is not necessarily large enough to be considered infinite and thus edge
effects could be present. In some measurements \cite{Decca} the sphere is located 
close to one end of the squared plane, to increase the torque exerted
on the underlying microresonator. Also, in other measurements a lens, schematized as a truncated 
sphere was used in lieu of a whole sphere \cite{Lamoreaux,prarc}. 
Fig.~\ref{fig:comsol_sp} shows the numerical results for configurations taking into 
account these deviations from the idealized case, as well as the curves expected 
from the exact (\ref{eq:sphereexact}) and the approximate (\ref{eq:sphereapprox}) expressions. 
For the whole sphere, it is worth remarking that neither the exact 
nor the approximate expression can give an accurate value for the capacitance 
between a sphere and a finite plane, even at small distance. We have checked that 
the idealized case of a sphere and an infinite plane is approached by considering 
square plates of progressively larger size. The capacitance in the realistic case 
still preserves a logarithmic dependence at small distance, with the slopes very close to each other. 
For the truncated sphere, the capacitance is significantly smaller than both 
the exact and the approximate expressions, which is expected considering that there is less 
conducting surface available in this case. Moreover, much smaller distance 
is required for the capacitance to be approximated by a logarithmic
dependence with a slope comparable to the one of the whole sphere case.  
The truncation aspect ratio has been chosen in close analogy on the
case of the lenses used in the sphere-plane experiment reported in
\cite{prarc}, and it should also be similar to the one used in the first modern Casimir 
force experiment in the sphere-plane configuration \cite{Lamoreaux}.

\section{4. Cylinder-plane geometry}

\begin{figure}[t]
\includegraphics[width=0.45\textwidth, clip=true]{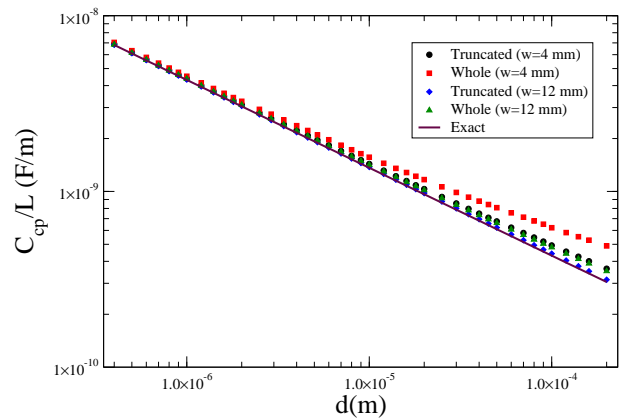}
\caption{Simulations in finite-size cylinder-plane configurations. 
Capacitance per unit of effective length {\it vs.} distance for a
truncated cylinder (black dots) and a whole 
cylinder (red squares) of width 4 mm in front of a square plane of size 10 mm $\times$ 28 mm, and 
similar plots for truncated (blue diamonds) and whole (green triangles) cylinders of width 12 mm. 
The radius of the cylinder is chosen to be 12 mm. 
The continuous line corresponds to the exact expression as 
in (\ref{eq:cylinderexact}), and the approximate expression (\ref{eq:cylinderapprox}) 
coincides with the exact expression in this distance range, confirming
that the PFA holds to a high degree of accuracy up to the largest
explored distances. }
\label{fig:comsol_cp}
\end{figure}

For the cylinder-plane geometry, the expression for the capacitance between 
a cylinder and an infinite plane is 
\begin{equation}
C_\mathrm{cp} = \frac{2 \pi \epsilon_0 L}{\cosh^{-1}(1 + d/R)}
\label{eq:cylinderexact}
\end{equation}
where $L$ and $R$ are the length and radius of the cylinder, and $d$ the distance.
In the limit of small separations, $d/R << 1$, an approximate expression for 
the capacitance can be obtained by expanding the inverse hyperbolic cosine 
function with Puiseux series \cite{Puiseux}:
\begin{equation}
C_\mathrm{cp} \approx \frac{\sqrt{2R} \pi \epsilon_0 L}{d^{0.5}}. 
\label{eq:cylinderapprox}
\end{equation}
Beside the finite size of the plane, the finite length of the
cylinder may also be a source of deviation from ideality. 
Therefore we have considered two cases, differing in the length of 
the cylinder chosen as smaller or larger than the side length of the
plane, and the analysis has been repeated for truncated cylinders. 
The capacitance per unit length $C_\mathrm{cp}/L_\mathrm{eff}$ 
{\it vs.} distance for different configurations are shown in Fig.~\ref{fig:comsol_cp}. 
In this figure $L_\mathrm{eff}$ is defined as the minimum between the length
of the cylinder and the length of the plane along the cylinder axis,
{\it i.e.} defining the {\sl overlapping} area between the cylinder and the plane. 
As can be seen in Fig.~\ref{fig:comsol_cp}, truncated, wider cylinders deviate 
less from the exact expression in comparison to whole, narrower
cylinders. However, longer width means harder parallelization and 
consequently worse achievable minimum distance. 
From Fig.~\ref{fig:comsol_cp}, a narrow truncated cylinder seems 
to be a good compromise, which is the configuration used in realistic experiments. 

\section{5. Implications for the electrostatic calibrations in Casimir
  force experiments} 

The discussion above shows that when the sizes of the samples are finite 
and the edge effects cannot be ignored, the capacitance will not scale 
with distance through the exponent expected from the ideal formula. 
If the exponent is incorrectly forced to the value from the ideal formula 
in the data fitting procedure, systematic errors will be propagated 
to all the fitting parameters. In fact, the optimal exponent is not a constant 
and depends on distance as evident in the plane-plane configuration shown 
in Fig.~\ref{fig:comsol_ppsquare}, therefore assuming a constant optimal 
exponent over the entire range of explored distances will generate 
systematic errors in the electrostatic calibrations. 

Let us consider the cylinder-plane geometry as an example. As already
mentioned, in a real experiment it is hard to determine directly the 
absolute separation distance, and only data fitting allows to obtain 
the separation if considered as a fitting parameter (see also the 
detailed discussion in \cite{Pala} on the same definition of distance
between two macroscopic bodies). In the ideal case, by fitting the 
capacitance data with a relationship like
\begin{equation}
C_\mathrm{cp} = C_0 + K_\mathrm{c}/(d-d_0)^{\epsilon_\mathrm{c}} 
\label{eq:cylinderfit}
\end{equation}
with the exponent $\epsilon_\mathrm{c}$ = 0.5 and $C_0$, 
$K_\mathrm{c}$ and $d_0$ as fitting parameters, 
the absolute distance can be obtained as $d_\mathrm{abs}=d-d_0$. 
Here $C_0$ represents a constant fitting parameter due to background, parasitic, 
distance-independent capacitance added to the cylinder-plane
capacitance, obtainable in the limit of very large gap separation. 
In the case of our numerical simulation data, we can fix $C_0=0$. 
If the correct fitting equation is used, 
the value of $d_0$ obtained should agree within the fitting errors with the expected 
null value. By fitting with Eq. (\ref{eq:cylinderfit}) the simulation data of the narrow 
truncated cylinder up to distance 10 $\mu$m, the maximum distance
at which electrostatic calibrations are usually carried out,
$d_0=-18\pm3$ nm is obtained with the exponent $\epsilon_\mathrm{c}$
is fixed at 0.5, while we obtain $d_0=12\pm1$ nm if $\epsilon_\mathrm{c}$ is left as a free parameter, 
with an optimal value of $0.4849\pm 0.0005$. In both cases, the value of $d_0$ from 
the fitting is different from zero by several standard deviations.
The smallest distance in the data is 400 nm, -18 nm and 12 nm
represent errors with respect to the smallest gap separation of 4.5 $\%$ and 3.0$\%$,
respectively. 
Similar analysis can be repeated for the parallel square plates, and the 
whole and truncated sphere configuration by fitting the data respectively with 
\begin{equation}
\begin{aligned}
C_\mathrm{pp} &= C_0 + K_\mathrm{c}/(d-d_0)^{\epsilon_\mathrm{c}}, \\
C_\mathrm{sp} &= C_0 + K_\mathrm{c} \mathrm{ln}(d-d_0).
\end{aligned}
\label{eq:ppspherefit}
\end{equation}
The results are shown in Table I, where the values of $d_0$ are all significantly different 
from the expected null value. To show that this is not a fitting artifact,
the same fitting procedures have been repeated with simulation data at
much smaller distance where the edge effects can be safely ignored. 
In this check the values of $d_0$ agree with the expected value of zero within 
the associated error from the fitting procedure. 
Therefore it is evident that the presence of edge effects may result in significant 
systematic errors in the best fit, affecting the accuracy of 
crucial parameters such as the absolute distance. 

With regard to Casimir experiments, other crucial parameters of the 
experiment beside the absolute distance between the two surfaces 
are determined through the electrostatic calibrations, for instance 
the effective mass $m_\mathrm{eff}$ of the resonator or, equivalently, its stiffness.
The calibrations are usually done at relatively large distance to ensure 
that the Casimir force can be neglected. However, in this regime the 
edge effects become more significant leading to systematic errors 
in the fitting procedures as shown above.
However, in most Casimir experiments what is measured in the
calibrations, rather than the capacitance, is the electrostatic 
force between the two surfaces or, if a resonator is used for 
dynamical measurements, its square frequency shift \cite{Bressicqg1}. 

The relationship between these two observables and the capacitance is in 
general given by 
\begin{equation}
F_\mathrm{el} = - \frac{\partial E_\mathrm{el}(d)}{\partial d} = 
-\frac{1}{2}\frac{\partial C(d)}{\partial d} V^2
\end{equation}
\begin{equation}
\Delta\nu^2_\mathrm{el} = -\frac{1}{4\pi^2 m_\mathrm{eff}} \frac{\partial F(d)}{\partial
  d} = \frac{1}{8 \pi^2 m_\mathrm{eff}}\frac{\partial^2 C(d)}{\partial d^2} V^2.
\end{equation}
Then the absolute value of the capacitance in itself is irrelevant 
for force or square frequency shift measurements and, due to the first 
and second derivatives, the distance dependence of the force or 
frequency shift signal may end up being less sensitive to the edge effects. 

\begin{table}[h]
\begin{center}
\begin{tabular}{|c|c|c|c|c|}
	\hline \hline
                        &   Parallel plates  & Sphere I & Sphere II   & Cylinder \\
	\hline
$d_0$(nm)               &  $-90\pm20$        & $13\pm3$ & $144\pm13$    &   $-18\pm3$  \\
$d_0$/$d_\mathrm{min}$  &  $1.8\%$           & $6.5\%$  & $72\%$        &   $4.5\%$    \\
	\hline
$d_0'$(nm)              &  $140\pm20$        & N/A & N/A          &   $12\pm1$   \\
$d_0'$/$d_\mathrm{min}$ &  $2.8\%$           & N/A & N/A          &   $3.0\%$    \\
$\epsilon_\mathrm{c}$   &  $0.977\pm0.002$   & N/A & N/A          &   $0.4849\pm0.0005$  \\
	\hline \hline    
\end{tabular}
\caption{Parameters of the best fit obtained for the capacitance data in the case of the four configurations 
considered in the text with Sphere I being whole sphere and Sphere II truncated sphere. 
For parallel plates and cylinder-plane
configurations, the top two rows are the parameters obtained when the exponent $\epsilon_\mathrm{c}$ is fixed at 
0.5 and 1, and the bottom three rows when $\epsilon_\mathrm{c}$ is left as a free
parameter. 
For the sphere-plane configuration, due to the logarithmic dependence
of the capacitance scales with distance, no exponent can be introduced.}
\end{center}
\label{tab:tablecap}
\end{table}

\begin{table*}[t]
\begin{center}
\begin{tabular}{|l|c|c|c|c|c|}

       \hline \hline
Geometry        & Fitting range ($\mu$m)     &  $K_\mathrm{f} (\mathrm{N m^{\epsilon_{f}}V^2}) $        &   $\epsilon_\mathrm{f}$       &   $d_0$(nm) 		& $d_0/d_\mathrm{min}$	\\
       \hline 
Parallel plates  & $7 - 50$       & $(3.59\pm 0.03)\times 10^{-16}$   &$1.9971 \pm 0.0007$  &$10.5 \pm 2.9$	& $0.15\%$	    \\
                 & $50 - 450$     &$(4.4 \pm 1.0)\times 10^{-16}$     &$1.973 \pm 0.025$    &$800 \pm 760$ 	& $1.6\%$  \\
       \hline 
Whole sphere     & $0.2 - 3$      &$(6.1 \pm 6.8)\times 10^{-15}$   &$0.97 \pm 0.08 $     &2$\pm$19  		& $1.0\%$ \\
	             & $3 - 25$       &$(1.4 \pm 2.0)\times 10^{-15}$   &$1.10 \pm 0.12$      &-430$\pm$410     & $14.3\%$ \\
	      \hline
Truncated sphere & $0.2 - 3$      &$(4.3 \pm 2.8)\times 10^{-17}$   &$1.30 \pm 0.04 $     &-35$\pm$10  		& $17.5\%$ \\
                 & $3 - 25$       &$(1.4 \pm 0.6)\times 10^{-21}$   &$1.92 \pm 0.04$      &-480$\pm$80   & $16.0\%$ \\
       \hline 
Truncated cylinder & $0.5 - 10$   &$(36.4\pm 2.2)\times 10^{-16}$ &$1.513\pm 0.005$     &$-5.8\pm 2.0$    &$1.2\%$ \\
                   & $10 - 160$   &$(29.2\pm 6.2)\times 10^{-16}$ &$1.538\pm 0.020$     &$-370\pm 180$    &$3.7\%$ \\

\hline
\hline
\end{tabular}
\caption{Parameters of the best fit obtained considering an
electrostatic-like force obtained leaving the power exponent for the distance dependence, 
as a free parameter in the cases of the four
configurations considered in the text. The parameter $K_\mathrm{f}$ is
the electrostatic coefficient, $\epsilon_\mathrm{f}$ the optimal
exponent for the force measurement, and $d_0$ an offset in distance, 
such that the electrostatic force can be expressed as 
$F_\mathrm{el}=F_0 + K_\mathrm{f}V^2/(d-d_0)^{\epsilon_{\mathrm{f}}}$.}
\end{center}
\label{tab:tableforce}
\end{table*}

To check the influence of the edge effects on force measurements, 
numerical differentiation has been carried out, using the Lagrange three-point 
formula \cite{Combrinck}, on the simulation capacitance data. The numerical 
differentiation method works better when the data points are equally
spaced, therefore in the case of the sphere-plane
configuration the numerical differentiation is performed in
the semilogarithmic domain, while for plane-plane and cylinder-plane 
configurations in the logarithmic domain. 
Then we fit the force {\it vs.} distance data with the corresponding 
equations for each geometry, obtaining the results shown in Table II. 
It seems that in terms of the optimal exponent, the edge effects are 
less significant when the force data are considered, apart from the case 
of the truncated sphere. More specifically, the results for the 
truncated sphere can be understood as if it behaves like a sphere 
only at small distance, instead resembles a plane when considered 
at large distance from the plane. This result is at variance 
with respect to the behavior actually observed in \cite{PRARC}, since 
a square frequency shift power law dependence on distance with 
an exponent less than 2 (therefore corresponding to a force exponent 
$\epsilon_\mathrm{f}<1$) has been observed uniformly over the entire range 
of explored distances. Therefore the origin of the anomaly observed in
\cite{PRARC} cannot be ascribed to edge effects. Analogous
considerations can be repeated for the first derivative of 
the force whenever square frequency shifts are measured.  
The influence of the edge effects on the values of fitting parameters $d_0$ and 
$K_\mathrm{f}$ in force measurement is also less significant at relatively small 
distance in the case of parallel plates, truncated cylinder and whole sphere, 
while in the case of the truncated sphere the error is much more significant. 
However, although the edge effects are not as strong in force measurement 
as in capacitance measurement, related systematic errors are still present, 
especially at large distance. Therefore the edge effects need to be considered 
and carefully examined if high precision measurements are required,
and the robustness of the fitting procedure with respect to the choice
of the distance range becomes crucial for the parameters, in
particular $K_\mathrm{f}$ and $d_0$. 

\section{6. Conclusion}

We have analyzed the effect of finite size in the predictions for the 
electrostatic calibrations of geometries of interest for measuring the Casimir force. 
One relevant consideration arising, or better quantitatively confirmed, from this 
numerical study is that there is a trade-off in the choice of the distance range 
to perform electrostatic calibrations. If this range is including too close distances, 
roughness corrections and the presence of the attractive Casimir force may result  
in systematic effects, while if large distance data are also considered one
must include edge effects in the theoretical modelization of the
experimental setup for a targeted level of accuracy of the electrostatic calibrations. 
If geometrical configurations manifestly differing from the idealized cases are 
also considered, such as a small portion of a sphere or a truncated cylinder, the 
deviations in the capacitance are even more pronounced and a full numerical evaluation 
of the electrostatics is necessary even at relatively moderate levels of desired accuracy. 
We have also shown that force (gradient force) measurements, based on the first (second) 
derivative with respect to distance of the capacitance, are in general more robust than 
absolute capacitance measurement in itself with respect to the geometrical deviations from 
ideality, confirming from another perspective a point recently discussed in \cite{Deccacapacitance}. 
Finally, we want to point out that detecting edge effects provides
also a simple and reliable method to assess the precision of the electrostatic
calibrations, for instance determining the minimum separation gap above which
deviations attributable to finite edges are effectively observed. 

\section{Acknowledgments}

We acknowledge useful discussions with D. A. R. Dalvit.

\end{document}